\documentclass[onecolumn,aps,amssymb,floatfix,prl]{revtex4}
\usepackage{amsmath}
\usepackage{amsfonts}
\usepackage{graphicx}

\begin{document}

\title{Description of self-synchronization effects in distributed Josephson
junction arrays using harmonic analysis and power balance}

\author{J. Hassel, L. Gr\"onberg, P. Helist\"o, H. Sepp\"a}
\address{VTT, P.O. Box 1000,
FIN-02044 VTT, Finland}

\begin{abstract}
Power generation and synchronisation in Josephson junction arrays have
attracted attention for a long time. This stems from fundamental interest in
nonlinear coupled systems as well as from potential in practical
applications. In this paper we study the case of an array of junctions
coupled to a distributed transmission line either driven by an external
microwave or in a self-oscillating mode.\ We simplify the theoretical
treatment in terms of harmonic analysis and power balance. We apply the
model to explain the large operation margins of SNS- and SINIS-junction
arrays. We show the validity of the approach by comparing with experiments
and simulations with self-oscillating es-SIS junction arrays.
\end{abstract}
\maketitle

Josephson junctions (JJs) are well known to be able to
generate microwave power. Their dynamics can also phase-lock either to an
external microwave bias or to the self-generated signal from an array of
junctions \cite{jai1}. A common configuration is to couple a number of JJs
through a distributed transmission line resonator \cite{wan1}. Both
overdamped \cite{wan1} \ and underdamped \cite{bar1} junctions in this
configuration\ are known to be able to mutually synchronize. Long arrays of
JJs coupled to a transmission line driven by external microwave bias are
known to exhibit properties, which can only be explained by the collective
dynamics of the JJ-transmission line system \cite{sch1,has1}.

Coupled Josephson junction arrays have been modelled by semiclassical
dynamical simulations \cite{caw1,alm1,fil1,kim1}, or in terms of lengthy
perturbation analysis \cite{tsy1}. The dynamics of coupled arrays is complex
making engineering of such devices challenging, often based on trial and
error. However, in most practical situations it suffices to study the
situation, where all the junctions are phase-locked to a driving signal
(either external or self-generated). Recognizing this, we develop a
simplified modelling technique. We have previously considered the case of
external microwave\ bias, for which we calculated the gain in the limit of
very small power self-generation \cite{has1}. Here we extend the analysis to
the case of arbitrarily large gain, and apply it to self-stabilized
overdamped JJ arrays. We analyze also the self-oscillating array by means of
a power balance equation. Finally, we compare the results of the analytic
model with simulations and experiments.

We assume that the tunnel element is effectively voltage biased.
This is ensured by a small impednace 
$\left| Z\left( f\right) \right| \ll \Phi _{0}f/I_{c}$, connected in 
parallel with the tunnel element (see Fig.
1(a)). Here $I_{c}$ is the critical current, $\Phi _{0}$ is the flux quantum and 
$f=\omega /2\pi$ is the drive frequency. We further assume that the 
dynamics is phase-locked to the 
driving signal.  
Below we drop the explicit $f$-dependence from the 
function notation, since all the quantities are
evaluated at that frequency. The condition for phase-locking (to the first
Shapiro step) is \cite{kau1}

\begin{equation}
\left| \delta I_{0}\right| <I_{c}\left| J_{1}\left( \widetilde{i}_{1}\right)
\right| ,  \label{stabcrit}
\end{equation}
where $\widetilde{i}_{1}=\left| Z\right| I_{1}/\Phi _{0}f$, $\delta
I_{0}=\left( I_{b}-\Phi _{0}f/R\right) $ and $J_{1}$ is the Bessel function
of the first kind. Here $I_{b}$ is the DC bias current, $R$ is the DC
resistance in parallel to the junction and $I_{1}$ is the amplitude of the
driving signal.

Using the assumptions above, we get the fundamental frequency component of
the Josephson current $I_{J1}$ (see Fig. 1(a)) \cite{has1}. Within harmonic
approximation the voltage across the tunnel element is 
$V_{J1}\left(I_{1}\right) =Z\left( I_{1}-I_{J1}\left( I_{1}\right) \right)$,
where the phase is referred to the driving signal $I_1$.
We define dynamical impedance 
$Z_{d}\left( I_{1}\right) =R_{d}+iX_{d}=V_{J1}\left(I_{1}\right) /I_{1}$, 
in series with 
$\omega L_{t}\gg X_{d}$, where 
$L_{t}$
 is the
transmission line inductance per junction length. We neglect 
$X_{d}$, 
assuming
$X_{d}< \omega L_{t}$. 
The effects discussed here follow from the real part 
$R_{d}\left( I_{1}\right) =\mathrm{Re}\left( Z_{d}\left( I_{1}\right) \right) =
\mathrm{Re}\left( Z\right) -\mathrm{Re}\left( ZI_{J1}/I_{1}\right) $. 
$\mathrm{Re}\left( Z\right)$ 
is the real part of the linear impedance parallel to the
junction, and $\mathrm{Re}( ZI_{J1}/I_1)$ is due to the Josephson effect. 

   \begin{figure}

    \includegraphics[width=7cm]{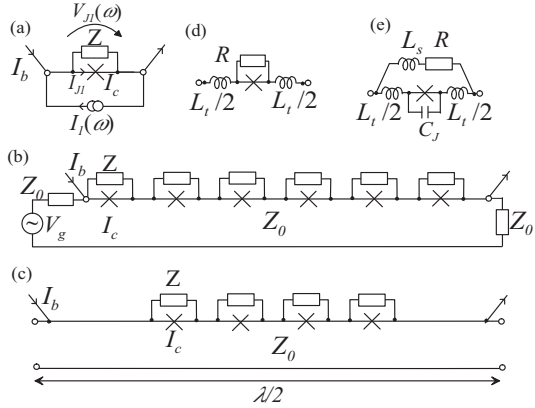}

    \caption{(a) A microwave current biased junction. (b) A Josephson array coupled to
a matched transmission line with external RF source. (c)\ A Josephson array
coupled to a $\lambda /2$-resonator. An $n\lambda /2$ resonator is a
straightforward generalization. (d) JJ with resistive environment. (e)\ JJ
with capacitive environment.}

    \end{figure}

Knowing $R_{d},$ various properties of arrays are readily obtained. We
consider first the case of external microwave drive (Fig. 1(b)). With $%
\left| R_{d}\right| \ll \omega L_{t}$ the (nonlinear) attenuation per
junction is

\begin{equation}
\eta \left( I_{1}\left( x_{k}\right) \right) =-\frac{1}{2}\frac{R_{d}\left(
I_{1}\left( x_{k}\right) \right) }{Z_{0}},  \label{tlgain}
\end{equation}
where $Z_{0}$ is the transmission line impedance. If \ $R_{d}>0$, the
junction attenuates and if $R_{d}<0$, it amplifies the driving current. This
is a generalization from the theory of low-loss transmission lines (e.g. 
\cite{poz1}). The current distribution $I_{1}\left( x_{k}\right) $ along an
array ($x_{k}$ being the position of the $k$:th junction)\ can be
recursively obtained from $I_{1}\left( x_{k+1}\right) =I_{1}\left(
x_{k}\right) \exp \left( \eta \left( I_{1}\left( x_{k}\right) \right)
\right) $. This also generalizes our previous results \cite{has1} to arbitrarily large
gains.

Our second case is a resonator driven by self-generated power of the JJ:s
(Fig. 1(c)). We note that the generated (or dissipated) power of a JJ at
position $x_{k}$ is $-R_{d}\left( I_{1}\left( x_{k}\right) \right)
I_{1}\left( x_{k}\right) ^{2}/2$. The current distribution in the system can
therefore be solved by means of a power balance equation 
\begin{equation}
-\frac{1}{2}\sum_{k=1}^{N}R_{d}\left( I_{1}\left( x_{k}\right) \right)
\times I_{1}\left( x_{k}\right) ^{2}=P_{ex},  \label{pobal0}
\end{equation}
where the left side is the generation due to all $N$ JJ:s and $P_{ex}$ is
the excess loss, i.e. all power loss mechanisms not related to the junction
or the shunt. In a high-Q resonator with $\left| R_{d}\right| ,\left|
X_{d}\right| \ll \omega L_{t}$ the resonant frequency determines the
generated frequency. The position dependence of $I_{1}\left( x\right) $
depends also only on the resonator properties, whereas the overall amplitude
depends on the gain and dissipation of JJ:s and on $P_{ex}$. The criterion (%
\ref{stabcrit}) needs to be satisfied for all $I_{1}\left( x_{k}\right) $.

We identify two limits for $R_{d}$ in terms of junction type. The first is a
resistive environment (Fig. 1(d)): $Z=R$ at the drive frequency. This is
relevant to SNS or SINIS junctions with appropriate parameters. In this case
we get, using above definition and results of Ref. \cite{has1} for $%
R_{d,R}=R_{d}\left( Z=R\right) $

\begin{equation}
R_{d,R}\left( I_{1}\right) =R\left( 1-2\frac{I_{c}R}{\Phi _{0}f}\frac{%
J_{1}^{\prime }\left( \widetilde{i_{1}}\right) }{\widetilde{i_{1}}}\sqrt{%
1-\left( \frac{\delta I_{0}}{I_{c}}\right) ^{2}J_{1}^{-2}\left( \widetilde{%
i_{1}}\right) }\right) ,  \label{odrd}
\end{equation}
in which $J_{1}^{\prime }$ is the derivative of the Bessel function.

The second limit is a capacitive environment, $Z\approx 1/j\omega C$, which
is valid for unshunted SIS\ junctions. The subgap leakage resistance can
typically be neglected. This is also a good approximation for externally
shunted SIS (es-SIS) junctions, if $\left| R+\omega L_{s}\right| \gg
1/\omega C$ (Fig. 1(e)). Here $L_{s}$ is the series inductance of the shunt
resistor. We get

\begin{equation}
R_{d,C}\left( I_{1}\right) =\mathrm{Re}\left( Z\right) -\frac{\Phi _{0}f}{2}%
\frac{\delta I_{0}}{I_{1}^{2}},  \label{udrd}
\end{equation}
where $\mathrm{Re}(Z) \approx 0 $ for unshunted junctions. For
es-SIS junctions the losses in the shunts can be the dominant mechanism of
linear dissipation, although their effect on JJ oscillation (second term in (%
\ref{udrd})) is small. For the arrangement of Fig. 1(e), $\mathrm{Re}\left(
Z\right) \approx \left( L_{eff}/L_{s}\right) ^{2}R$ if $L_{s}\gg L_{eff}$,
where $L_{eff}=L_{t}-1/\omega ^{2}C$.

It is experimentally known that long overdamped arrays driven by external
microwave compensate for the transmission line attenuation \cite{sch1}. This
is the reason why SNS\ and SINIS Josephson voltage arrays work in spite of
the huge attenuation due to intrinsic damping \cite{sch1,ham1}. With the
above notation, self-compensation occurs if the amplitude of the propagating
signal is initially between $I_{1,\min }$\ and $I_{1,\max }$ such that (see (%
\ref{tlgain}), (\ref{odrd})) 
\begin{eqnarray}
\eta \left( I_{1}\right)  &=&0,I_{1}=I_{1}^{\prime },  \notag \\
\eta \left( I_{1}\right)  &>&0,I_{1,\min }<I_{1}<I_{1}^{\prime },
\label{stabcrit3} \\
\eta \left( I_{1}\right)  &<&0,I_{1}^{\prime }<I_{1}<I_{1,\max }.  \notag
\end{eqnarray}
In addition, $I_{1,\min }$\ and $I_{1,\max }$ need to satisfy (\ref{stabcrit}%
). Then $I_{1}\left( x\right) $ approaches the stable point $I_{1}^{\prime }$
towards the end of the transmission line. Fig. 2 shows the maximum step
width $2\delta I_{0,\max }$ and pump current margin $I_{1,\max }-I_{1,\min }$%
, within which the array self-stabilizes the microwave current. Typical
values of the design parameter $\Phi _{0}f/I_{c}R$ of overdamped arrays vary
between 0.5 and 2. In this range, the maximum self-stabilized step width
varies between 0.2$I_{c}$ and 0.6$I_{c}$. Using the parameters of Fig. 2 in 
\cite{sch1} ($\Phi _{0}f/I_{c}R\simeq 1.3$), our model predicts $2\delta
I_{0,\max }\approx 400$ $\mu $A and $20\log _{10}\left( I_{1,\max
}/I_{1,\min }\right) \approx 18$ dB, which are in excellent agreement with
the exprimental data ($2\delta I_{0,\max }\approx 400$ $\mu $A and $20\log
_{10}\left( I_{1,\max }/I_{1,\min }\right) \gtrsim 10$ dB).

   \begin{figure}

    \includegraphics[width=7cm]{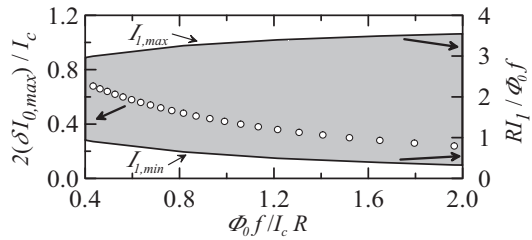}

    \caption{Operating margins of an overdamped JJ array in self-stabilizing mode 
according to (\ref{stabcrit}) and (\ref{stabcrit3}). Open circles (left axis): 
Maximum step width as a function of 
$\Phi _{0}f/I_{c}R$. Gray-shaded area (right axis): pump current range 
maximising the step width.}

    \end{figure}

In SIS and es-SIS junctions, $\eta $ does not change sign as function of $%
I_{1}$ (see the second term in (\ref{udrd})). Therefore the tendency for
power self-stabilization is not present. The much smaller attenuation of the
capacitive environment makes phase-locking of all junctions possible in this
case for a reasonable number of junctions \cite{has1,has2}.

Finally, to study self-oscillating arrays, we prepared arrays embedded in
open-ended $\lambda /2$ or $\lambda $ microstrip resonators. The circuits
were fabricated using the Nb/Al/AlOx/Nb fabrication line of VTT \cite{gro1}
and measured at 4.2 K. The resonators were 800 $\mu $m or 1600 $\mu $m long
Nb strips having a resonant frequency $f$ nominally about 80 GHz. The
resonators had 2-28 es-SIS JJ:s centered around the node(s) of the standing
wave. A photograph of one of the structures is shown in Fig. 3(a). Measured
IV curves are shown in Fig. 3(b).

The AC current distribution in the resonators is given as $I_{1}\left(
x\right) =I_{0}\sin \left( 2\pi x/\lambda \right) $. The excess loss is
given as $P_{ex}=\omega L_{T0}I_{0}^{2}/2Q.$ where $L_{T0}$ is the total
inductance of the resonator and $Q$ is the intrinsic Q-value. Using Eqs. (%
\ref{pobal0}),(\ref{udrd}) the power balance stands $\frac{1}{2}%
\sum_{k=1}^{N}\left( \left( L_{eff}/L_{s}\right) ^{2}R\times \left(
I_{0}\sin \left( 2\pi x_{k}/\lambda \right) \right) ^{2}-2\Phi _{0}f\times
\delta I_{0}\right) =\pi fL_{t0}I_{0}^{2}/Q$. This is solved for $I_{0}$.
For $\delta I_{0}<0,$ no physical solution exist, since $R_{d,C}$ can be
negative only for positive $\delta I_{0}$. We get

\begin{eqnarray}
I_{1}\left( x\right) &=&\sqrt{\frac{N\Phi _{0}f\times \delta I_{0}}{2\pi
fL_{t0}/2Q+\left( 1/2\right) \left( L_{eff}/L_{s}\right)
^{2}R\sum_{k=1}^{N}\sin ^{2}\left( 2\pi x_{k}/\lambda \right) }}\sin \left( 
\frac{2\pi x}{\lambda }\right) .  \label{pobal} \\
\delta I_{0} &<&I_{c}\left| J_{1}\left( \frac{I_{1}\left( x_{k}\right) }{%
2\pi \Phi _{0}f^{2}C}\right) \right| \text{.}  \label{stabcrit2}
\end{eqnarray}
Here we have rewritten (\ref{stabcrit}) for convenience. For the resistive
environment also the negative side of the step $\delta I_{0}<0$ appears.

The solutions for which all junctions are phase-locked to the driving signal
appear in the IV curves of Fig. 3 as steps with voltage $%
\left\langle V\right\rangle =N\Phi _{0}f$ (for simplicity we omit the small
dependence of $f$ on the bias point due to $X_{d}$). This happens for the
solutions of (\ref{pobal}) that satisfy (\ref{stabcrit2}) for all $x_{k}$.
Solutions where no phase locking occurs are described by a resistive IV
curve determined by the shunt resistors. We assume here that these are the
only types of solutions.

   \begin{figure}

    \includegraphics[width=7cm]{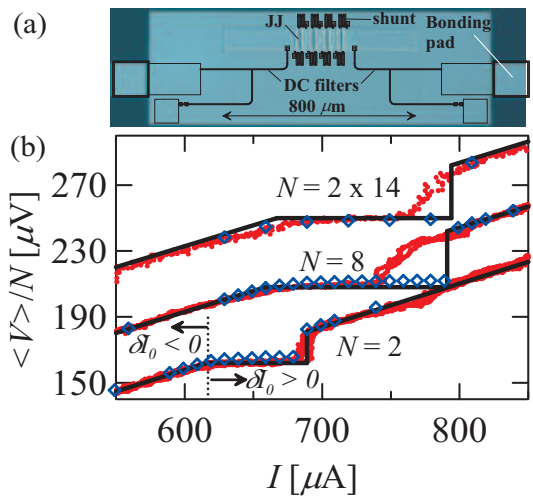}

    \caption{(a) An experimental array with a $\lambda /2$-resonator
having $N=8$ JJ:s. Typical circuit parameters are 
$I_{c}=280$ $\mu $A, $C=59$ pF, $R=0.26$ $\Omega $, $L_{s}=3.4$ pH, $%
L_{t}=0.2$ pH, $L_{t0}=5.7$ pH and $f=80$ GHz. (b) Current-voltage curves
measured from the resonators with $N$ JJs (small red circles). The device
with $N=2\times 14$ JJ:s is based on a $\lambda $-resonator. Others are
based on $\lambda /2$ resonators. The voltage is scaled to that of a single
junction and curves are lifted by 40 $\mu $V with respect to the one below
for clarity. The analytic curves are shown by solid black lines and the
simulated points are denoted by large blue diamonds.}

    \end{figure}

The theoretical IV curves obtained in this way are shown as solid lines in
Fig. 3(b). The fitting parameters are the Q-value and the resonant frequency 
$f$. The Q-value was fitted to the case with $N=2$, for which the effect of
the parameter spread is the smallest. The result $Q=420$ was used for all
the arrays. This corresponds to a transmission line loss of $\alpha =\pi
/\lambda Q\times 8.7$ dB $\approx 41$ dB/m, very close to that of an earlier
measurement (50 dB/m) \cite{has2}. The loss is dominated by the dielectric
(PECVD\ SiO$_{2}$) loss. The resonance frequencies for all devices were very
close to the nominal value 80 GHz. Other parameters were obtained
independently either from the IV curves or from the known circuit and
material properties.

The fact that steps appear only for $\delta I_{0}\gtrsim 0$ as expected is
clarified for $N=2$ in Fig. 3(b). The step widths are close to the values
predicted by the theory, though smaller. The difference is explained by
parameter spread (the shunt resistances varied up to $\pm 5\%$ within a
chip, changing the effective bias point $\delta I_{0}$ by tens of $\mu $A).

To further study the validity of the model we performed time-domain
simulations using Stewart-McCumber model for the junctions and a distributed
model of a dissipative transmission line \cite{poz1} to describe the
resonator sections. We found good agreement with the analytic model. The
model step widths are well reproduced, though the small bending of the
plateau due to finite $X_{d}$ is not accounted for by the single-frequency
model. As assumed, in all cases the numerical solutions were such that
either all the junctions were synchronized or nonsynchronized. However, for
simulated devices having even larger number of JJs per resonant node than our
experimental devices, we
found also partially synchronized states. Also in these cases the bias range
of the completely synchronized state was correctly predicted by the analytic
model.

In summary, we have developed a simple analytic technique for quantitative
modelling of distributed JJ\ systems with a strong preferred frequency set
by either an external RF signal or a high-Q resonance. The results clarify
differences between underdamped and overdamped systems, a source of some
controversy recently. The model provides an intuitive way of understanding
the dynamics of phase locking\ and power generation in JJ arrays. The model
predicts results in agreement with simulations and experiments and it can be
used to engineer practical devices, such as mm-wave local oscillators and JJ
voltage standards. An interesting alternative voltage standards is a JJ\
array with no external microwave source, locking the self-generated signal
directly to a frequency reference. The power balance equation is easily
generalized to externally loaded systems to calculate the power output.
Other types of physical systems can be analyzed similarly. For example, the
exact dual for systems described here is a distributed parallel array of
quantum phase slip junctions \cite{moo1}, where the ''conductor gain'' of JJ
arrays is replaced by a ''dielectric gain''.

The work was partially supported by MIKES\ (Centre for Metrology and
Accreditation), Academy of Finland (Centre of Excellence in Low Temperature
Quantum Phenomena and Devices) and EU through RSFQubit (FP6-3749). The
authors wish to thank Antti Manninen for useful discussions.


\begin{thebibliography}{99}
\bibitem{jai1}  A.\ K. Jain, K. K. Likahrev, J. E. Lukens, and J. E.
Sauvageau, Phys. Rep. 109, 310 (1984).

\bibitem{wan1}  K. Wan, A.K. Jain, and J.E. Lukens, Appl. Phys. Lett. 54,
1805 (1989).

\bibitem{bar1}  P. Barbara, A. B. Cawthorne, S. V. Shitov, and C.J. Lobb,
Phys. Rev. Lett. 82, 1963 (1999).

\bibitem{sch1}  H. Schulze, R. Behr, F. Mueller, and J. Niemeyer, Appl.
Phys. Lett. 73, 996, 1998.

\bibitem{has1}  J. Hassel, P. Helist\"{o}, L. Gr\"{o}nberg, H. Sepp\"{a}, J.
Nissil\"{a} and A. Kemppinen, IEEE Trans. Instrum. Meas. 54, 632 (2005).

\bibitem{kau1}  R. L. Kauz, and R. Monaco, J. Appl. Phys. 57, 875 (1985).

\bibitem{caw1}  A.\ B. Cawthorne, P. Barbara, S. V. Shitov, C. J. Lobb, K.
Wiesenfield, and A. Zangwill, Phys. Rev. B 60 (1999).

\bibitem{alm1}  E. Almaas, and D. Stroud, Phys. Rev. B. 65, 134502 (2002).

\bibitem{fil1}  G. Filatrella, N. F. Pedersen, C. J. Lobb, and P. Barbara,
Eur. Phys. J. B 34, 3 (2003).

\bibitem{kim1}  K.-T. Kim, M.-S.-Kim, Y. Chong, and J. Niemeyer, Appl. Phys.
Lett. 88, 062501 (2006).

\bibitem{tsy1}  D. Tsygankov, and K. Wiesenfeld, Phys. Rev. E 66, 036215
(2002).

\bibitem{poz1}  D. M. Pozar, ''Microwave Engineering'', 2nd. ed., John Wiley
\& Sons, inc. (1988).

\bibitem{ham1}  C.\ A. Hamilton, C. J. Burroughs, S. P. Benz, and J. R.
Kinard, IEEE Trans. Instrum. Meas. 46, 224 (1997).

\bibitem{has2}  J. Hassel, H. Sepp\"{a}, L. Gr\"{o}nberg, and I. Suni, Rev.
Sci. Instrum. 74, 3510 (2003).

\bibitem{gro1}  L. Gr\"{o}nberg, J. Hassel, P. Helist\"{o}, M. Kiviranta, H.
Sepp\"{a}, M. Kulawski, T. Riekkinen, and M. Ylilammi, Extended abstracts of
International Superconductivity Conference 2005, Noordwijkerhout, The
Netherlands, O-W.04 (2005).

\bibitem{moo1}  J. E. Mooij, and Yu. V. Nazarov, Nature Physics 2, 169-172
(2006).
\end{thebibliography}
\end{document}